\documentclass[10pt,a4paper,twocolumn,english,prd,showpacs]{revtex4}
\usepackage[T1]{fontenc}
\usepackage[latin9]{inputenc}
\usepackage{amsmath}
\usepackage{amssymb}
\usepackage{graphicx}
\usepackage{esint}

\makeatletter

\pdfpageheight\paperheight
\pdfpagewidth\paperwidth

\@ifundefined{textcolor}{}
{%
 \definecolor{BLACK}{gray}{0}
 \definecolor{WHITE}{gray}{1}
 \definecolor{RED}{rgb}{1,0,0}
 \definecolor{GREEN}{rgb}{0,1,0}
 \definecolor{BLUE}{rgb}{0,0,1}
 \definecolor{CYAN}{cmyk}{1,0,0,0}
 \definecolor{MAGENTA}{cmyk}{0,1,0,0}
 \definecolor{YELLOW}{cmyk}{0,0,1,0}
}

\usepackage{babel}

\usepackage{babel}

\makeatother

\usepackage{babel}
\begin{document}

\title{Chaotic dynamics of the Bianchi IX universe in Gauss-Bonnet gravity}

\author{Edward J. Kim}

\email{edward@skku.edu}

\selectlanguage{english}%

\affiliation{Department of Physics, Sungkyunkwan University, Suwon 440-746, Korea}

\author{Shinsuke Kawai}

\email{kawai@skku.edu}

\selectlanguage{english}%

\affiliation{Department of Physics, Sungkyunkwan University, Suwon 440-746, Korea}
\begin{abstract}
We investigate the dynamics of closed FRW universe and anisotropic
Bianchi type-IX universe characterized by two scale factors in a gravity
theory including a higher curvature (Gauss-Bonnet) term. The presence
of the cosmological constant creates a critical point of saddle type
in the phase space of the system. An orbit starting from a neighborhood
of the separatrix will evolve toward the critical point, and it eventually
either expands to the de Sitter space or collapses to the big crunch.
In the closed FRW model, the dynamics is reduced to hyperbolic motions
in the two-dimensional center manifold, and the system is not chaotic.
In the anisotropic model, anisotropy introduces the rotational mode,
which interacts with the hyperbolic mode to present a cylindrical
structure of unstable periodic orbits in the neighborhood of the critical
point. Due to the non-integrability of the system, the interaction
of rotational and hyperbolic modes makes the system chaotic, making
it impossible for us to predict the final fate of the universe. We
find that the chaotic dynamics arises from the fact that orbits with
even small perturbations around the separatrix oscillate in the neighborhood
of the critical point before finally expanding or collapsing. The
chaotic character is also evidenced by the fractal structures in the
basins of attraction.
\end{abstract}



\maketitle

\section{Introduction}

The pioneering work of Belinskii, Khalatnikov, and Lifshitz \cite{BKL1969}
has shown that anisotropic Bianchi type-IX universes could exhibit
chaotic dynamics as it approaches the cosmological singularity $\left(t\rightarrow0\right)$.
The anisotropic Bianchi IX universe, as it approaches the singularity,
exhibits an oscillatory mode consisting of an infinite sequence of
Kasner eras, during which two scale factors oscillate and the remaining
scale factor decreases monotonically. Independently, Misner \cite{misner1969mixmaster}
also suggested a chaotic approach to initial singularity in his Mixmaster
universe (vacuum Bianchi IX universe characterized by three scale
factors).

While the aspect of chaotic dynamics helped deepen our understanding
of singularities in general relativity, there have been significant
challenges in finding an invariant measure of chaos. A standard measure
of chaos is the Lyapunov exponent, and a debate started on whether
or not the Mixmaster universe is chaotic, when some studies found
zero Lyapunov exponent for the Mixmaster universe, while others found
a positive Lyapunov exponent \cite{francisco1988qualitative,burd1991chaos,berger1991comments}.
The conflict was resolved when it was realized \cite{Rugh1990} that
Lyapunov exponent is coordinate dependent and therefore is not a reliable
measure of chaos in general relativity.

Cornish and Levin \cite{cornish1996chaos} demonstrated the appearance
of chaos in the dynamics of the Friedmann-Robertson-Walker (FRW) models
with a cosmological constant and scalar fields conformally coupled
to the geometry. The presence of the cosmological constant creates
a saddle point in the phase space, and the separatrix connects this
saddle point to other critical points. When there are interactions
between the geometry with the scalar fields, the separatrix breaks
up and becomes fractal. The work by de Oliveira, Soares, and Stuchi
\cite{deOliveira1997chaos} emphasized that even small perturbations
induces the breaking of highly unstable separatrix. The presence of
a positive cosmological constant and a perfect fluid creates a saddle-center
in the phase space, and non-integrability of the system induces distortion
and twisting of the topology of homoclinic cylinders in the neighborhood
of the critical point. The fractal and topological structures are
coordinate-independent and thus provide invariant characterization
of chaos in relativistic theories.

Regarding such a system as a model of the early universe, a natural
question that arises is whether the Einstein gravity is reliable or
not. It has been pointed out \cite{Brandenberger:2012aj} that we
might have the trans-Planckian issue; at early stages of inflation,
the quantum gravity effects are not necessarily negligible. Previous
works on chaotic dynamics in the per-inflationary era focused on Einstein
gravity, but as we approach the Planck scale, it is reasonable to
expect higher order corrections to the Einstein-Hilbert action. Superstring
theory is the leading candidate for a description of physics at such
scale. Our goal in this paper is to use an invariant characterization
of chaos and investigate the existence of chaotic dynamics in ``string-inspired''
modified gravity. Our model takes into account the effects of the
higher curvature terms that typically arise in the one-loop low-energy
effective superstring action. For simplicity we focus on a model in
which a scalar field representing the string moduli is non-minimally
coupled to the Gauss-Bonnet curvature. For a related work on chaotic
dynamics of higher curvature modified gravity, see e.g. \cite{cotsakis1993mixmaster}. 

The Bianchi type-IX models with a scalar field coupled to the Gauss-Bonnet
term are studied in somewhat different context of singularity avoidance
in \cite{yajima2000generality}. The new features introduced by our
model are a positive cosmological constant and a perfect fluid, which
create saddle points in the phase space of the system. Because this
method does not work for the case of zero cosmological constant, our
work implies that in order for the universe to inflate we need to
include a positive cosmological constant in the low-energy effective
action. We introduce this modified action in Sec. \ref{sec:action}.
We first consider the closed FRW universe in Sec. \ref{sec:FRW},
where the degrees of freedom are the scale factor and a scalar field,
and discuss the basic characteristics of the closed FRW model. We
introduce anisotropy in the metric in Sec. \ref{sec:Axisymmetric-Bianchi-IX}
and discuss how anisotropy creates a topological structure of cylinders
near the critical point. In Sec. \ref{sec:numerical} we present numerical
evidence of cylindrical topology, oscillatory behavior around the
critical point, and fractal structures in the basins of attraction
when small perturbations are introduced in the metric and/or the scalar
field. The fractal structures in the basins of attraction, as well
as the topology of cylinders, constitute invariant characterization
of chaos, and we conclude that chaotic dynamics can arise in our string-inspired
model.

\section{One-loop effective action\label{sec:action}}

We start with the action in the Einstein frame given by \cite{antoniadis1994topological,rizos1994existence,easther1996one,kawai1998instability,kawai1999nonsingular,yajima2000generality}
\begin{multline}
S=\int\mathrm{d}^{4}x\sqrt{-g}\left[\frac{1}{2}R-\Lambda-\frac{1}{2}\left(\nabla\sigma\right)^{2}-\frac{\lambda}{16}\xi\left(\sigma\right)R_{GB}^{2}\right]\\
+S_{matter}\label{eq:action}
\end{multline}
where $R$ and $\sigma$ are the Ricci scalar curvature and a scalar
field, respectively. In our units, the gravitational constant corresponds
to $G=1/8\pi$. A positive cosmological constant is necessary for
the existence of a saddle point in phase space and the universe to
inflate. The Gauss-Bonnet curvature is given by 
\begin{equation}
R_{GB}^{2}=R^{2}-4R^{\alpha\beta}R_{\alpha\beta}+R^{\alpha\beta\gamma\delta}R_{\alpha\beta\gamma\delta}\label{eq:gauss_bonnet}
\end{equation}
and the function $\xi\left(\sigma\right)$ depends on details of string
theory compactified geometry \cite{antoniadis1992moduli}. For example,
when $\sigma$ is the dilaton, we have 
\begin{equation}
\xi\left(\sigma\right)=e^{\sigma}
\end{equation}
In type II superstring, $\sigma$ can be regarded as the modulus of
compactified dimensions. The form of $\xi\left(\sigma\right)$ in
a particular compactification can be found in \cite{antoniadis1992moduli,antoniadis1994topological}.
The function $\xi\left(\sigma\right)$ determines the coupling of
$\sigma$ and the geometry, and is expressed with Dedekind $\eta$
function as 
\begin{equation}
\xi\left(\sigma\right)=-\ln\left[2e^{\sigma}\eta^{4}\left(ie^{\sigma}\right)\right]\label{eq:dedekind_function}
\end{equation}
Due to the modular property 
\begin{equation}
\eta\left(-\frac{1}{\tau}\right)=\sqrt{-i\tau}\eta\left(\tau\right)
\end{equation}
the function $\xi\left(\sigma\right)$ is even in $\sigma$, has a
global minimum at $\sigma=0$, and increases exponentially as $\sigma\rightarrow\pm\infty$.
We will be interested in the small $\sigma$ behavior in the following
sections. We assume for simplicity that $\xi\left(\sigma\right)$
can be approximated as 
\begin{equation}
\xi\left(\sigma\right)=\frac{1}{2}\sigma^{2}
\end{equation}

\section{Closed Friedmann-Robertson-Walker universe \label{sec:FRW}}

We first consider a cosmological model characterized by the scale
factor $a\left(t\right)$ with the line element given by 
\begin{equation}
\mathrm{d}s^{2}=-\mathrm{d}t^{2}+a^{2}\left(t\right)\left[\left(\omega^{1}\right)^{2}+\left(\omega^{2}\right)^{2}+\left(\omega^{3}\right)^{2}\right]
\end{equation}
The Bianchi type-IX invariant one-forms $\omega^{i}$ are given by
\begin{align}
\omega^{1} & =\sin\psi\,\mathrm{d}\theta-\sin\theta\cos\psi\,\mathrm{d}\phi\nonumber \\
\omega^{2} & =\cos\psi\,\mathrm{d}\theta+\sin\theta\sin\psi\,\mathrm{d}\phi\nonumber \\
\omega^{3} & =\cos\theta\,\mathrm{d}\phi+\mathrm{d}\psi
\end{align}
which are chosen to satisfy $\mathrm{d}\omega^{i}=-\frac{1}{2}{\epsilon^{i}}_{jk}\omega^{j}\wedge\omega^{k}$
and ${\epsilon^{i}}_{jk}$ is the completely antisymmetric tensor
$\left({\epsilon^{1}}_{23}=1\right)$. This represents the closed
FRW universe with positive curvature $k=+1$. We assume that the matter
content is a perfect fluid. The energy-momentum tensor of the perfect
fluid can be written in the form 
\begin{equation}
T^{\mu\nu}=\left(\rho+p\right)\delta_{0}^{\mu}\delta_{0}^{\nu}+pg^{\mu\nu}\label{eq:energy_momentum_tensor_perfect_fluid}
\end{equation}
where $\rho$ and $p$ are energy density and pressure, respectively.
For simplicity, we assume that the perfect fluid may be represented
by ``dust,'' that is, $\gamma=0$ in the equation of state $p=\gamma\rho$.
However, even for general perfect fluids, we can expect similar features
\cite{barguine2001homoclinic}.

For the action \eqref{eq:action} the Lagrangian is given by 
\begin{equation}
L=\frac{3}{4}a-\Lambda a^{3}-3a\dot{a}^{2}+\frac{1}{2}a^{3}\dot{\sigma}^{2}+\frac{\lambda}{8}\sigma\dot{\sigma}\left(3\dot{a}+4\dot{a}^{3}\right)
\end{equation}
where the overdot denotes differentiation with respect to time $t$.
The equations of motion are 
\begin{multline*}
-\frac{1}{4}+\Lambda a^{2}-\dot{a}^{2}-\frac{1}{2}a^{2}\dot{\sigma}^{2}-2a\ddot{a}\\
+\frac{1}{8}\lambda\left(\dot{\sigma}^{2}+4\dot{a}^{2}\dot{\sigma}^{2}+8\sigma\dot{\sigma}\dot{a}\ddot{a}+\sigma\ddot{\sigma}+4\sigma\ddot{\sigma}\dot{a}^{2}\right)=0
\end{multline*}
\begin{equation}
3a^{2}\dot{a}\dot{\sigma}+a^{3}\ddot{\sigma}+\frac{3}{8}\lambda\sigma\ddot{a}\left(1+4\dot{a}^{2}\right)=0\label{eq:eom}
\end{equation}
with the Hamiltonian constraint given by 
\begin{equation}
H=-3a\dot{a}^{2}-\frac{3}{4}a+a^{3}\Lambda+\frac{3}{8}\lambda\sigma\dot{\sigma}\dot{a}\left(1+4\dot{a}^{2}\right)+\frac{1}{2}a^{3}\dot{\sigma}^{2}+E_{0}=0
\end{equation}
where $E_{0}=\rho a^{3}$ corresponds to the total matter content
of the model.

We can always decompose a set of second-order differential equations
into a set of first-order differential equations by redefining variables.
Thus, we get a set of four coupled first-order ordinary differential
equations 
\[
\frac{\mathrm{d}}{\mathrm{d}t}a=\dot{a},\qquad\frac{\mathrm{d}}{\mathrm{d}t}\sigma=\dot{\sigma}
\]
\begin{multline*}
\frac{\mathrm{d}}{\mathrm{d}t}\dot{a}=8a^{2}\bigg[-3\lambda\sigma\dot{\sigma}\dot{a}\left(1+4\dot{a}^{2}\right)\\
+a^{3}\left(8\Lambda-4\dot{\sigma}^{2}\right)+a\left(1+4\dot{a}^{2}\right)\left(-2+\lambda\dot{\sigma}^{2}\right)\bigg]\chi
\end{multline*}
\begin{multline}
\frac{\mathrm{d}}{\mathrm{d}t}\dot{\sigma}=\Bigg\{-384a^{3}\dot{a}\dot{\sigma}+3\lambda\sigma\bigg(-\left(1+4\dot{a}^{2}\right)^{2}\left(-2+\lambda\dot{\sigma}^{2}\right)\\
+4a^{2}\left[-2\Lambda+\dot{\sigma}^{2}+4\dot{a}^{2}\left(-2\Lambda+5\dot{\sigma}^{2}\right)\right]\bigg)\Bigg\}\chi\label{eq:eom_first_order}
\end{multline}
where $\chi$ is given by 
\begin{equation}
\chi=\frac{1}{64a^{3}\left(2a-\lambda\dot{a}\sigma\dot{\sigma}\right)+3\lambda^{2}\sigma^{2}\left(1+4\dot{a}^{2}\right)^{2}}
\end{equation}
The dynamical system \eqref{eq:eom_first_order} admits the static
Einstein universe as a solution, which is just the critical point
$P$. Its coordinates are given by 
\begin{align}
P:\; a & =a_{0}\equiv\frac{1}{2\sqrt{\Lambda}},\;\sigma=\sigma_{0},\;\dot{a}=0,\,\dot{\sigma}=0
\end{align}
where $\sigma_{0}$ is a constant. The critical energy is given by
\begin{equation}
E_{crit}=\frac{1}{4\sqrt{\Lambda}}
\end{equation}

When the scalar field is absent and does not interact with the curvature,
the equations of motion can be integrated exactly, and the dynamical
system admits an invariant manifold $\mathcal{M}$. The invariant
manifold $\mathcal{M}$ is defined by 
\begin{equation}
\mathcal{M}:\;\sigma=0,\quad\dot{\sigma}=0
\end{equation}
and the dynamical system simplifies to an exactly integrable two-dimensional
system 
\[
\frac{\mathrm{d}}{\mathrm{d}t}a=\dot{a}
\]
\begin{equation}
\frac{\mathrm{d}}{\mathrm{d}t}\dot{a}=-\frac{1}{8a}-\frac{\dot{a}^{2}}{2a}+\frac{1}{2}\Lambda a\label{eq:invariant_manifold_eom}
\end{equation}

\begin{figure}
\noindent \begin{centering}
\includegraphics[scale=0.5]{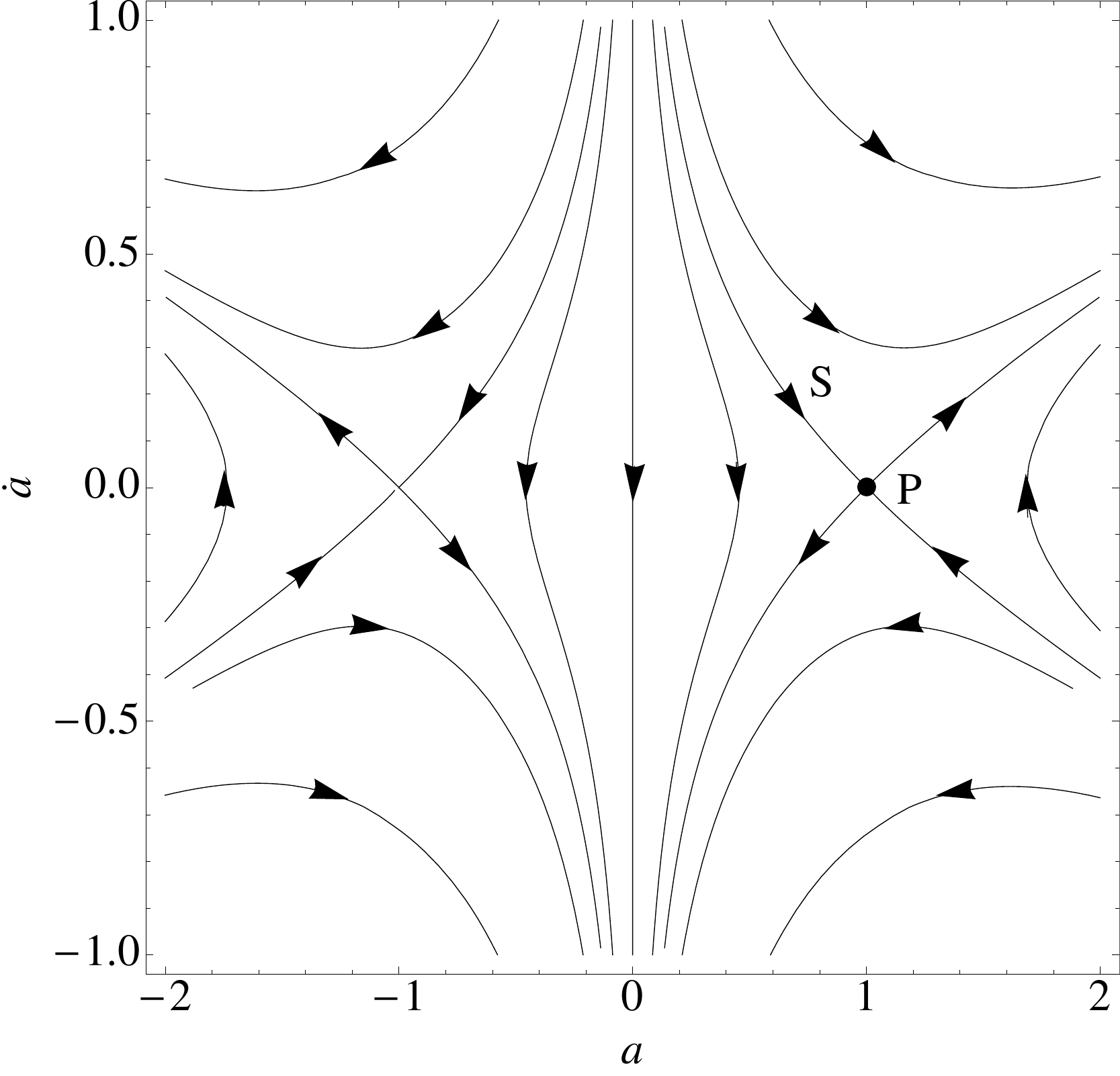} 
\par\end{centering}

\caption{Phase space portrait of the invariant manifold $\mathcal{M}$ in conformal
time $\mathrm{d}\eta=\mathrm{d}t/a$. The orbit $S$ is the separatrix
and the point $P$ is the critical point.}

\label{fig:invariant_manifold_sigma} 
\end{figure}

In Fig. \ref{fig:invariant_manifold_sigma} we display the phase space
portrait of the invariant manifold $\mathcal{M}$ in the $\left(a,\dot{a}\right)$
plane. We introduce the conformal time $\mathrm{d}\eta=\mathrm{d}t/a$
so that the dynamical system \eqref{eq:invariant_manifold_eom} becomes
regular at $a=0$. The phase space is divided by a separatrix into
two types of orbits: those that collapse into the big crunch and those
that expand into the de Sitter space. The critical point $P$ intersects
the invariant manifold $\mathcal{M}$ at $\sigma_{0}=0$. Furthermore,
there are attractors corresponding to stable and unstable de Sitter
spaces. When the dynamics is dominated by the cosmological constant
$\Lambda$, it can be shown that the scale factor $a\left(t\right)$
expands exponentially and approaches the stable de Sitter attractor
as $a\left(t\right)\sim e^{\sqrt{\Lambda/3}t}$.

Let us now linearize the dynamical equations \eqref{eq:eom_first_order}
about the critical point $P$. We move the critical point to the origin
by redefining 
\begin{align}
a\rightarrow a+a_{0},\quad & \sigma\rightarrow\sigma+0\nonumber \\
\dot{a}\rightarrow\dot{a}+0,\quad & \dot{\sigma}\rightarrow\dot{\sigma}+0
\end{align}
then we obtain 
\begin{equation}
\frac{\mathrm{d}}{\mathrm{d}t}\begin{pmatrix}a\\
\sigma\\
\dot{a}\\
\dot{\sigma}
\end{pmatrix}=A_{0}\begin{pmatrix}a\\
\sigma\\
\dot{a}\\
\dot{\sigma}
\end{pmatrix}+\begin{pmatrix}\text{higher}\\
\text{order}\\
\text{terms}
\end{pmatrix}\label{eq:eom_origin}
\end{equation}
where the constant matrix associated with linearizing the system \eqref{eq:eom_first_order}
about the critical point $P$ is given by 
\begin{equation}
A_{0}=\begin{pmatrix}0 & 0 & 1 & 0\\
0 & 0 & 0 & 1\\
\Lambda & 0 & 0 & 0\\
0 & 0 & 0 & 0
\end{pmatrix}\label{eq:matrix_linearized}
\end{equation}
The matrix \eqref{eq:matrix_linearized} has four eigenvalues 
\begin{equation}
\lambda_{1,2}=\pm\sqrt{\Lambda},\quad\lambda_{3,4}=0^{2}\label{eq:eigenvalues_frw}
\end{equation}

According to the center manifold theorem \cite{wiggins2003introduction},
the study of the dynamics near the critical point can be reduced to
the study of the dynamics restricted to the associated two-dimensional
invariant manifolds $W^{c}\left(0\right)$ near the critical point.
Without actually calculating the center manifold, we argue as follows.
Under the coordinate transformation 
\begin{equation}
\begin{pmatrix}a\\
\sigma\\
\dot{a}\\
\dot{\sigma}
\end{pmatrix}=\begin{pmatrix}0 & 0 & -\frac{1}{\sqrt{\Lambda}} & \frac{1}{\sqrt{\Lambda}}\\
1 & 0 & 0 & 0\\
0 & 0 & 1 & 1\\
0 & 1 & 0 & 0
\end{pmatrix}\begin{pmatrix}u_{1}\\
u_{2}\\
u_{3}\\
u_{4}
\end{pmatrix}
\end{equation}
the matrix \eqref{eq:matrix_linearized} assumes the Jordan canonical
form 
\begin{equation}
J=\begin{pmatrix}0 & 1 & 0 & 0\\
0 & 0 & 0 & 0\\
0 & 0 & -\sqrt{\Lambda} & 0\\
0 & 0 & 0 & \sqrt{\Lambda}
\end{pmatrix}
\end{equation}
Thus the four-dimensional system \eqref{eq:eom_first_order} restricted
on the center manifold $W^{c}\left(0\right)$ is a two-dimensional
vector field whose linear part is given by 
\begin{equation}
\begin{pmatrix}0 & 1\\
0 & 0
\end{pmatrix}\label{eq:bogdanov_linear_part}
\end{equation}
Bogdanov \cite{bogdanov1975versal} has shown that the system which
has a linear part \eqref{eq:bogdanov_linear_part} is locally topologically
equivalent near the critical point to the normal form 
\[
\dot{v}_{1}=v_{2}
\]
\begin{eqnarray}
\dot{v}_{2} & = & \beta_{1}+\beta_{2}v_{2}+v_{1}^{2}+\gamma v_{1}v_{2}\label{eq:bogdanov_normal_form}
\end{eqnarray}
where $\gamma=\pm1$. The dynamics of \eqref{eq:bogdanov_normal_form}
is not qualitatively changed by the higher order terms in the normal
form. The normal form \eqref{eq:bogdanov_normal_form} has no homoclinic
or periodic solutions for $\beta_{1}>0$. If we rescale the variables
and parameters as 
\begin{equation}
v_{1}=\epsilon^{2}q_{1},\quad v_{2}=\epsilon^{3}p_{1},\quad\beta_{1}=\epsilon^{4}\mu,\quad\beta_{2}=\epsilon^{2}\label{eq:bogdanov_rescale}
\end{equation}
where $\epsilon>0$ and rescale the time as 
\begin{equation}
t\rightarrow\frac{t}{\epsilon}
\end{equation}
then the system becomes 
\[
\dot{q}_{1}=p_{1}
\]
\begin{equation}
\dot{p}_{1}=\mu+q_{1}^{2}+\epsilon\left(p_{1}+\gamma q_{1}p_{1}\right)\label{eq:eom_normal_form}
\end{equation}
For $\epsilon=0$, the rescaled equations become exactly integrable
Hamiltonian system with the Hamiltonian given by 
\begin{equation}
H\left(q_{1},p_{1}\right)=\frac{p_{1}^{2}}{2}+\frac{q_{1}^{3}}{3}+\mu q_{1}\label{eq:hamiltonian_normal_form}
\end{equation}
where $\mu$ is a parameter with $\mu>0$ corresponding to the near-zero
eigenvalues being real. For $\mu<0$ there exist homoclinic solutions.
However, we restrict our attention to only the hyperbolic motions
and do not consider the homoclinic orbits, because homoclinic orbits
extend to $a<0$ region of the phase space, which does not have a
physical meaning.

In the parlance of nonlinear dynamical systems theory, this is the
Bogdanov-Takens bifurcation. To perform a global analysis that includes
the effect of the $\mathcal{O}\left(\epsilon\right)$ part of \eqref{eq:eom_normal_form}
on this integrable structure, Melinkov's method \cite{wiggins2003introduction}
can be used. A detailed description of bifurcation diagrams and phase
portraits of the Bogdanov-Takens bifurcation can be found elsewhere
(cf. \cite{arnold1988geometricalmethods}) and need not be repeated
here.

What the above linear analysis shows is that although we could expect
simple hyperbolic motion of the planar Bogdanov-Takens bifurcation,
we do not expect chaotic dynamics in our closed FRW model. As we will
see in Sec. \ref{sec:Axisymmetric-Bianchi-IX}, the chaotic dynamics,
especially the rotational mode of the oscillatory Mixmaster dynamics,
arises from the presence of imaginary eigenvalues of the linearized
matrix.

\section{Axisymmetric Bianchi IX Universe \label{sec:Axisymmetric-Bianchi-IX}}

Chaos is expected to come from the anisotropy of the universe. For
simplicity, we restrict our attention to two distinct scale factors
$a\left(t\right)$ and $b\left(t\right)$ and consider an axisymmetric
Bianchi IX cosmological model with the line element 
\begin{equation}
ds^{2}=-dt^{2}+a^{2}\left(t\right)\left[\left(\omega^{1}\right)^{2}+\left(\omega^{2}\right)^{2}\right]+b^{2}\left(t\right)\left(\omega^{3}\right)^{2}\label{eq:metric_axisymmetric}
\end{equation}
For the action \eqref{eq:action} the Lagrangian is given by 
\begin{multline}
L=b-\Lambda a^{2}b-\frac{b^{3}}{4a^{2}}-b\dot{a}^{2}-2a\dot{a}\dot{b}+\frac{1}{2}a^{2}b\dot{\sigma}^{2}\\
+\frac{\lambda}{8}\sigma\dot{\sigma}\left(2\frac{b^{3}\dot{a}}{a^{3}}+4\dot{b}-3\frac{b^{2}\dot{b}}{a^{2}}+4\dot{a}^{2}\dot{b}\right)
\end{multline}
and the equations of motion are 
\begin{multline*}
2\Lambda ab-\frac{b^{3}}{2a^{3}}-2\dot{a}\dot{b}-ab\dot{\sigma}^{2}-2\ddot{a}b-2a\ddot{b}\\
+\lambda\left[\left(\frac{b^{3}}{4a^{3}}+\dot{a}\dot{b}\right)\left(\dot{\sigma}^{2}+\sigma\ddot{\sigma}\right)+\sigma\dot{\sigma}\left(\dot{a}\ddot{b}+\ddot{a}\dot{b}\right)\right]=0
\end{multline*}
\begin{multline*}
-1+\Lambda a^{2}+\frac{3b^{2}}{4a^{2}}-\dot{a}^{2}-\frac{1}{2}a^{2}\dot{\sigma}^{2}-2a\ddot{a}\\
+\lambda\left[\left(\frac{1}{2}+\frac{\dot{a}^{2}}{2}-\frac{3b^{2}}{8a^{2}}\right)\left(\dot{\sigma}^{2}+\sigma\ddot{\sigma}\right)+\dot{a}\ddot{a}\sigma\dot{\sigma}\right]=0
\end{multline*}
\begin{multline}
2ab\dot{a}\dot{\sigma}+a^{2}\dot{b}\dot{\sigma}+a^{2}b\ddot{\sigma}\\
+\frac{\lambda}{8}\sigma\bigg[-6\frac{b^{3}\dot{a}^{2}}{a^{4}}+12\frac{b^{2}\dot{a}\dot{b}}{a^{3}}-6\frac{b\dot{b}^{2}}{a^{2}}+2\frac{b^{3}\ddot{a}}{a^{3}}\\
+8\dot{a}\dot{b}\ddot{a}+4\ddot{b}-3\frac{b^{2}\ddot{b}}{a^{2}}+4\dot{a}^{2}\ddot{b}\bigg]=0\label{eq:eom_axisymmetric}
\end{multline}
with the Hamiltonian constraint given by 
\begin{multline}
H=-\dot{a}^{2}b-2a\dot{a}\dot{b}-b+\frac{b^{3}}{4a^{2}}+a^{2}b\Lambda+\frac{1}{2}a^{2}b\dot{\sigma}^{2}\\
+\frac{\lambda}{8}\sigma\dot{\sigma}\left(2\frac{\dot{a}b^{3}}{a^{3}}+4\dot{b}-3\frac{b^{2}\dot{b}}{a^{2}}+12\dot{a}^{2}\dot{b}\right)+E_{0}=0
\end{multline}
where $E_{0}=\rho a^{2}b$ corresponds to the total matter content
of the model.

We can always decompose a set of second-order differential equations
into a set of first-order differential equations by redefining variables.
Thus we get a set of six coupled first-order differential equations
in the variables $\left(a,b,\sigma,\dot{a},\dot{b},\dot{\sigma}\right)$.
The dynamical system \eqref{eq:eom_axisymmetric} admits the static
Einstein universe as a solution, which is just the critical point
$P$. Its coordinates are given by 
\begin{equation}
P:\; a=b=a_{0}\equiv\frac{1}{2\sqrt{\Lambda}},\;\sigma=\sigma_{0},\;\dot{a}=\dot{b}=\dot{\sigma}=0
\end{equation}
where $\sigma_{0}$ is a constant. The critical energy is given by
\begin{equation}
E_{crit}=\frac{1}{4\sqrt{\Lambda}}
\end{equation}

When the scalar field is absent and the isotropy is restored, the
equations of motion are exactly integrable and the dynamical system
admits the same invariant manifold $\mathcal{M}$ of Fig. \ref{fig:invariant_manifold_sigma}.
As before, the critical point $P$ intersects the invariant manifold
$\mathcal{M}$ at $\sigma_{0}=0$.

The constant matrix associated with linearizing the system \eqref{eq:eom_axisymmetric}
about the critical point $P$ is given by 
\begin{equation}
A_{0}=\begin{pmatrix}0 & 0 & 0 & 1 & 0 & 0\\
0 & 0 & 0 & 0 & 1 & 0\\
0 & 0 & 0 & 0 & 0 & 1\\
-2\Lambda & 3\Lambda & 0 & 0 & 0 & 0\\
6\Lambda & -5\Lambda & 0 & 0 & 0 & 0\\
0 & 0 & 0 & 0 & 0 & 0
\end{pmatrix}\label{eq:matrix_linearized_anisotropic}
\end{equation}
The matrix \eqref{eq:matrix_linearized_anisotropic} has six eigenvalues
\begin{equation}
\lambda_{1,2}=\pm\sqrt{\Lambda},\quad\lambda_{3,4}=0^{2},\quad\lambda_{5,6}=\pm2i\sqrt{2\Lambda}
\end{equation}
Compared to the four eigenvalues \eqref{eq:eigenvalues_frw} in the
isotropic case, the anisotropy in the metric has produced an additional
pair of imaginary eigenvalues. Without loss of generality, we fix
$\Lambda=1/4$ so that $a_{0}=1$. Under the transformation 
\begin{equation}
\begin{pmatrix}a\\
b\\
\sigma\\
\dot{a}\\
\dot{b}\\
\dot{\sigma}
\end{pmatrix}=\begin{pmatrix}0 & 0 & -2 & 2 & \frac{1}{2\sqrt{2}} & -\frac{1}{2\sqrt{2}}\\
0 & 0 & -2 & 2 & -\frac{1}{\sqrt{2}} & \frac{1}{\sqrt{2}}\\
1 & 0 & 0 & 0 & 0 & 0\\
0 & 0 & 1 & 1 & -\frac{1}{2} & -\frac{1}{2}\\
0 & 0 & 1 & 1 & 1 & 1\\
0 & 1 & 0 & 0 & 0 & 0
\end{pmatrix}\begin{pmatrix}q_{1}\\
p_{1}\\
q_{2}\\
p_{2}\\
q_{3}\\
p_{3}
\end{pmatrix}
\end{equation}
the matrix \eqref{eq:matrix_linearized_anisotropic} assumes the Jordan
canonical form 
\begin{equation}
J=\begin{pmatrix}0 & 1 & 0 & 0 & 0 & 0\\
0 & 0 & 0 & 0 & 0 & 0\\
0 & 0 & -\frac{1}{2} & 0 & 0 & 0\\
0 & 0 & 0 & \frac{1}{2} & 0 & 0\\
0 & 0 & 0 & 0 & 0 & -\sqrt{2}\\
0 & 0 & 0 & 0 & \sqrt{2} & 0
\end{pmatrix}
\end{equation}
Thus, with canonical transformation, the Hamiltonian to the quadratic
order can be expressed as 
\begin{multline}
H\sim-\frac{1}{4}\left(p_{2}^{2}-q_{2}^{2}\right)+p_{1}^{2}+\frac{1}{\sqrt{2}}\left(p_{3}^{2}+q_{3}^{2}\right)+E_{0}-E_{crit}\\
+\mathcal{O}\left(3\right)=0
\end{multline}

In a small neighborhood of the critical point, the higher order terms
in the expansion $\mathcal{O}\left(3\right)$ is negligible and we
may assume that the energy $E_{0}-E_{crit}$ is small. Then the Hamiltonian
may be approximated as 
\begin{equation}
H\sim-\frac{1}{4}\left(p_{2}^{2}-q_{2}^{2}\right)+p_{1}^{2}+\frac{1}{\sqrt{2}}\left(p_{3}^{2}+q_{3}^{2}\right)+E_{0}-E_{crit}=0\label{eq:hamiltonian_approximate}
\end{equation}
In this linear regime, the Hamiltonian \eqref{eq:hamiltonian_approximate}
is separable. If we define the partial energies as 
\begin{equation}
E_{1}=p_{1}^{2}
\end{equation}
\begin{equation}
E_{hyp}=\frac{1}{4}\left(p_{2}^{2}-q_{2}^{2}\right)
\end{equation}
\begin{equation}
E_{rot}=\frac{1}{\sqrt{2}}\left(p_{3}^{2}+q_{3}^{2}\right)
\end{equation}
they are approximately conserved separately, 
\begin{equation}
-E_{hyp}+E_{rot}+E_{1}=E_{crit}-E_{0}
\end{equation}

The topological structure near the critical point created by these
separable partial energies in the case of general relativity was first
described by de Oliveira, Soares, and Stuchi in \cite{deOliveira1997chaos}.
Compared to their model, we have an additional partial energy $E_{1}$.
Ignoring the energy $E_{1}$ for the moment, let us concentrate on
the hyperbolic motion energy $E_{hyp}$ and the rotational motion
energy $E_{rot}$. If $E_{hyp}=0$, we have either $p_{2}=q_{2}=0$
or $p_{2}=\pm q_{2}$. When $p_{2}=q_{2}=0$, we have $p_{3}^{2}+q_{3}^{2}=\text{constant}$
and the motion will be described by periodic orbits $\tau_{E_{0}}$
in the $\left(p_{1},q_{1}\right)$ plane. These periodic orbits depend
on the parameter $E_{0}$. When $p_{2}=\pm q_{2}$, the motion will
be described by linear stable $V_{s}$ and unstable $V_{u}$ one-dimensional
manifolds in the $\left(p_{2},q_{2}\right)$ plane. In addition, we
have $E_{1}=p_{1}^{2}=\text{constant}$, which generates a linear
one-dimensional manifold $\Gamma_{E_{0}}$. The direct product of
$\tau_{E_{0}}$ with $V_{s}$, $V_{u}$, and $\Gamma_{E_{0}}$ generates
the topological structure of stable cylinders $\tau_{E_{0}}\times\Gamma_{E_{0}}\times V_{s}$
and unstable cylinders $\tau_{E_{0}}\times\Gamma_{E_{0}}\times V_{u}$.
Thus the flow in the phase space will be $\mathbb{R}^{2}\times S^{1}$.

The center manifold is the nonlinear extension of the linear regime
that corresponds to $E_{hyp}=0$. The intersection of this center
manifold with the energy surface $E_{0}=E_{crit}$ corresponds to
$p_{1}=p_{2}=q_{2}=0$, which is just the critical point $P$. Since
$E_{rot}$ and $E_{1}$ is always positive, for $E_{0}>E_{crit}$
the center manifold does not intersect the energy surface in the linear
regime. However, when we include nonlinear terms, the Gauss-Bonnet
term may have positive as well as negative energy. Therefore, contrary
to the general relativity case, the center manifold can intersect
the energy surface even if $E_{0}>E_{crit}$, and we may have a saddle
structure for $E_{0}>E_{crit}$.

In general, an orbit that approaches the neighborhood of the critical
point $P$ will have $E_{1}\neq0$, $E_{hyp}\neq0$, and $E_{rot}\neq0$.
As this orbit with energy $E_{crit}-E_{0}$ approaches the critical
point, the non-integrability of the system \eqref{eq:eom_axisymmetric}
makes it impossible to predict the amount of energy that will be partitioned
into each mode $E_{1}$, $E_{hyp}$, or $E_{rot}$. Since we cannot
predict how much energy will be transformed into which mode, we cannot
predict whether the orbit will collapse into the big crunch or escape
to inflation. The rotational mode arising from anisotropy corresponds
to the oscillatory mode of the Mixmaster dynamics and is crucial in
this non-predictability.

In the parlance of nonlinear dynamical systems theory, the double
zero eigenvalues lead to Bogdanov-Takens bifurcation and a pair of
purely complex eigenvalues lead to Andronov-Hopf bifurcation \cite{wiggins2003introduction}
in the four-dimensional center manifold. We have new possibilities
in our system due to the interaction of the Bogdanov-Takens bifurcation
with the Andronov-Hopf bifurcation. The simple hyperbolic motion of
the planar Bogdanov-Takens bifurcation transversally intersects two-dimensional
stable and unstable manifolds of periodic orbits, and this interaction
leads to chaotic dynamics.

\section{Numerical Results\label{sec:numerical}}

In our numerical experiments, we closely follow the approach of de
Oliveria, Soares, and Stuchi \cite{deOliveira1997chaos}. We fix $\Lambda=0.25$
so that the coordinates of the critical point $P$ are given by $a=b=1.0$,
$\sigma=0$, $\dot{a}=\dot{b}=\dot{\sigma}=0$, and $E_{crit}=0.5$.
Let $S_{0}$ be a point on the separatrix $S$ in the invariant manifold
$\mathcal{M}$ $\left(\sigma=\dot{\sigma}=0\right)$. For the results
presented in this paper, we choose the coordinates $a_{0}=b_{0}=0.9$
and $\dot{a}_{0}=\dot{b}_{0}=0.051\,818\,772\,5$, but any point near
the separatrix would yield similar results. Around $S_{0}$, we perturb
the separatrix in five variables (three in the closed FRW case) by
an arbitrarily small amount $\delta=10^{-4}$ and use the Hamiltonian
constraint to fix the remaining variable. The energy of the orbit
$E_{0}$ is chosen to be very close to the energy of the separatrix
so that the difference in energy $\Delta E_{0}=E_{crit}-E_{0}$ is
much smaller than the perturbation $\delta$. This mimics the uncertainty
in the initial conditions. In physical terms, these initial conditions
represent small perturbations in the scale factor and/or the scalar
field in the pre-inflationary era.

\begin{figure}
\includegraphics[scale=0.5]{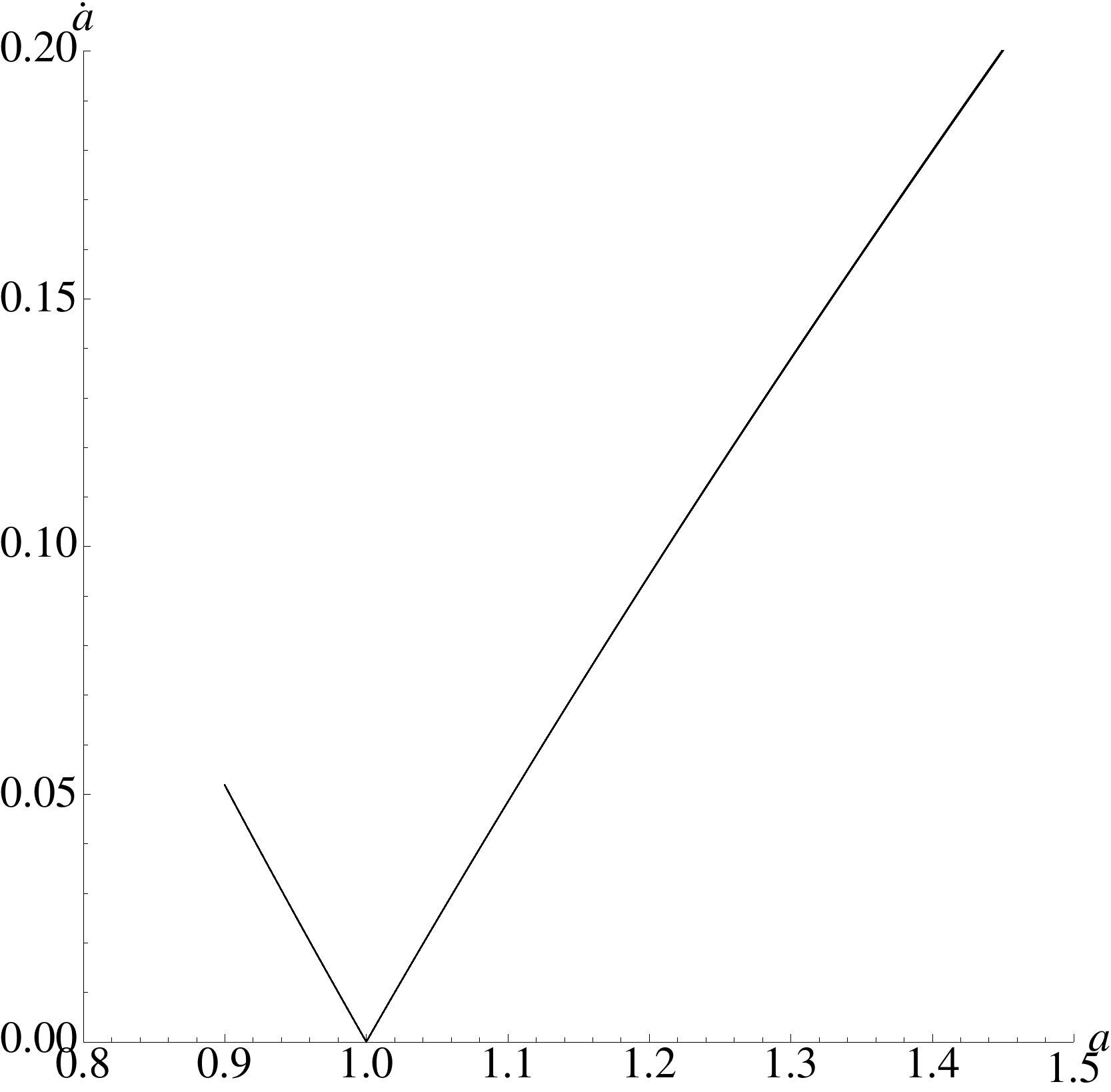} 

\caption{Escape to inflation of 100 orbits around a point on the separatrix
with coordinates $a=b=0.9$, $\dot{a}=\dot{b}=0.051\,818\,772\,5$,
and $\sigma=\dot{\sigma}=0$. The coupling constant is $\lambda=16$.
The energy surface is given by $E_{0}=0.499\,999\,999\,9$ and the
radius of the sphere of initial conditions is $\delta=10^{-4}$. \label{fig:outline_inflation}}
\end{figure}

\begin{figure}
\includegraphics[scale=0.5]{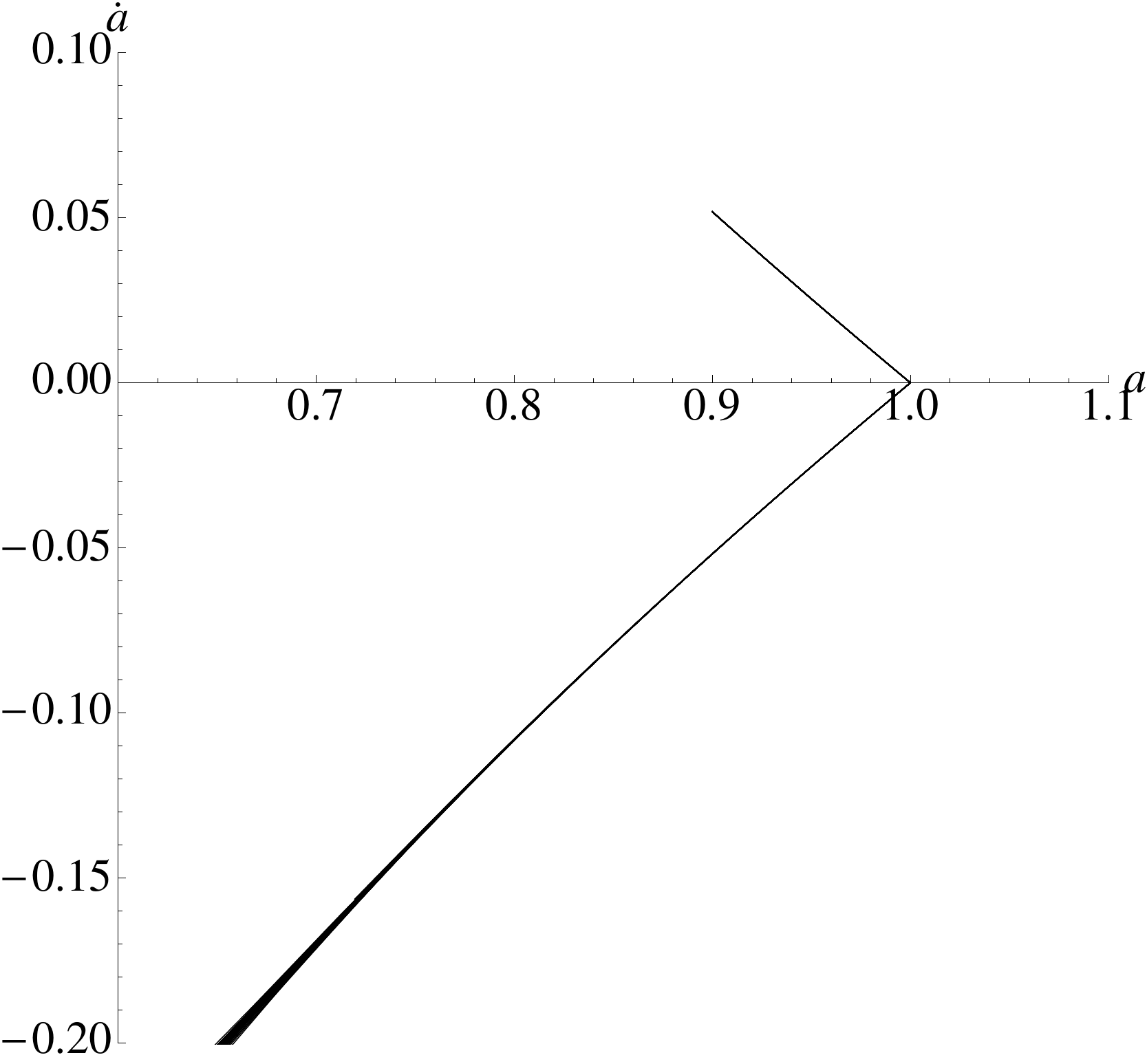}

\caption{Collapse of 100 orbits around the same point in Fig. \ref{fig:outline_inflation}
with $\lambda=16$, $E_{0}=0.499\,999\,980\,2$, and $\delta=10^{-4}$.\label{fig:outline_collapse}}
\end{figure}

When the orbits are evolved starting from the point $S_{0}$, the
orbits evolve toward the critical point since the initial conditions
were chosen near the separatrix. After passing through the critical
point, we expect the orbits to either collapse into the big crunch
or expand to de Sitter space. The two possible outcomes for 100 orbits
with $\delta=10^{-4}$ are shown for the inflation in Fig. \ref{fig:outline_inflation}
and for the collapse in Fig. \ref{fig:outline_collapse}. The final
fate of the orbits depends on the energy $E_{0}$, and there exists
an upper bound on $E_{0}$ for which all orbits collapse and a lower
bound on $E_{0}$ for which all orbits escape to inflation. For an
energy between this interval, some orbits collapse and other orbits
inflate, resulting in an indeterminate outcome.

\begin{figure}
\includegraphics[scale=0.5]{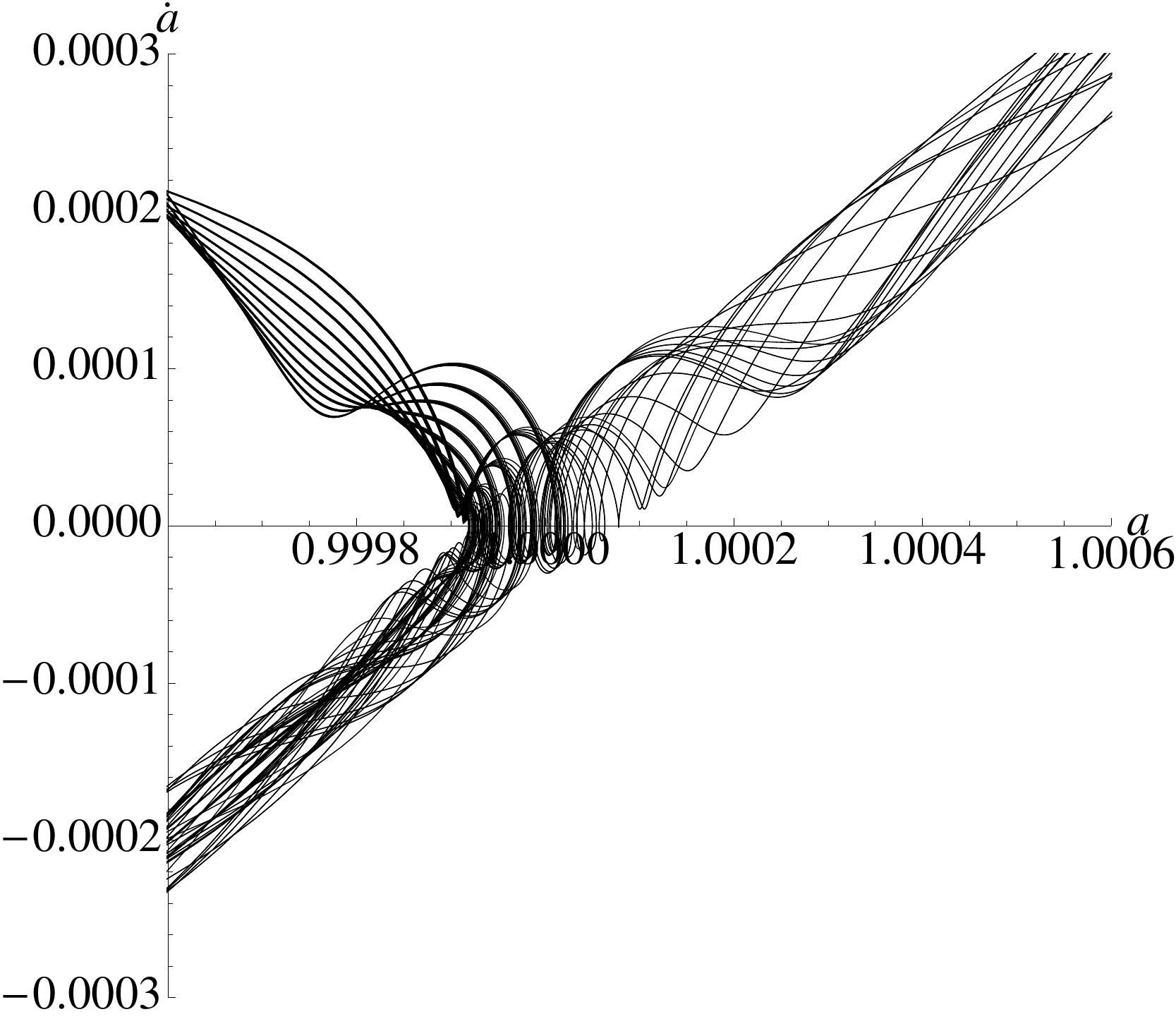}

\caption{A magnified view of 100 orbits around the critical point in the $\left(a,\dot{a}\right)$
plane for $\delta=10^{-4}$ and $E_{0}=0.499\,999\,997\,5$.\label{fig:magnified_critical_point}}
\end{figure}

In Fig. \ref{fig:magnified_critical_point} we display a magnified
view of the region around the critical point in the $\left(a,\dot{a}\right)$
plane. Note that the orbits oscillate around the separatrix as well
as the critical point. This oscillatory mode is crucial in the mixing
of the boundaries and the existence of chaotic dynamics. As this orbit
with energy $E_{crit}-E_{0}$ approaches the critical point, the non-integrability
of the system makes it impossible to predict the amount of energy
that will be partitioned into each mode $E_{1}$, $E_{hyp}$, or $E_{rot}$.
Thus, the outcome has a sensitive dependence on initial conditions.

\begin{figure}
\includegraphics[scale=0.5]{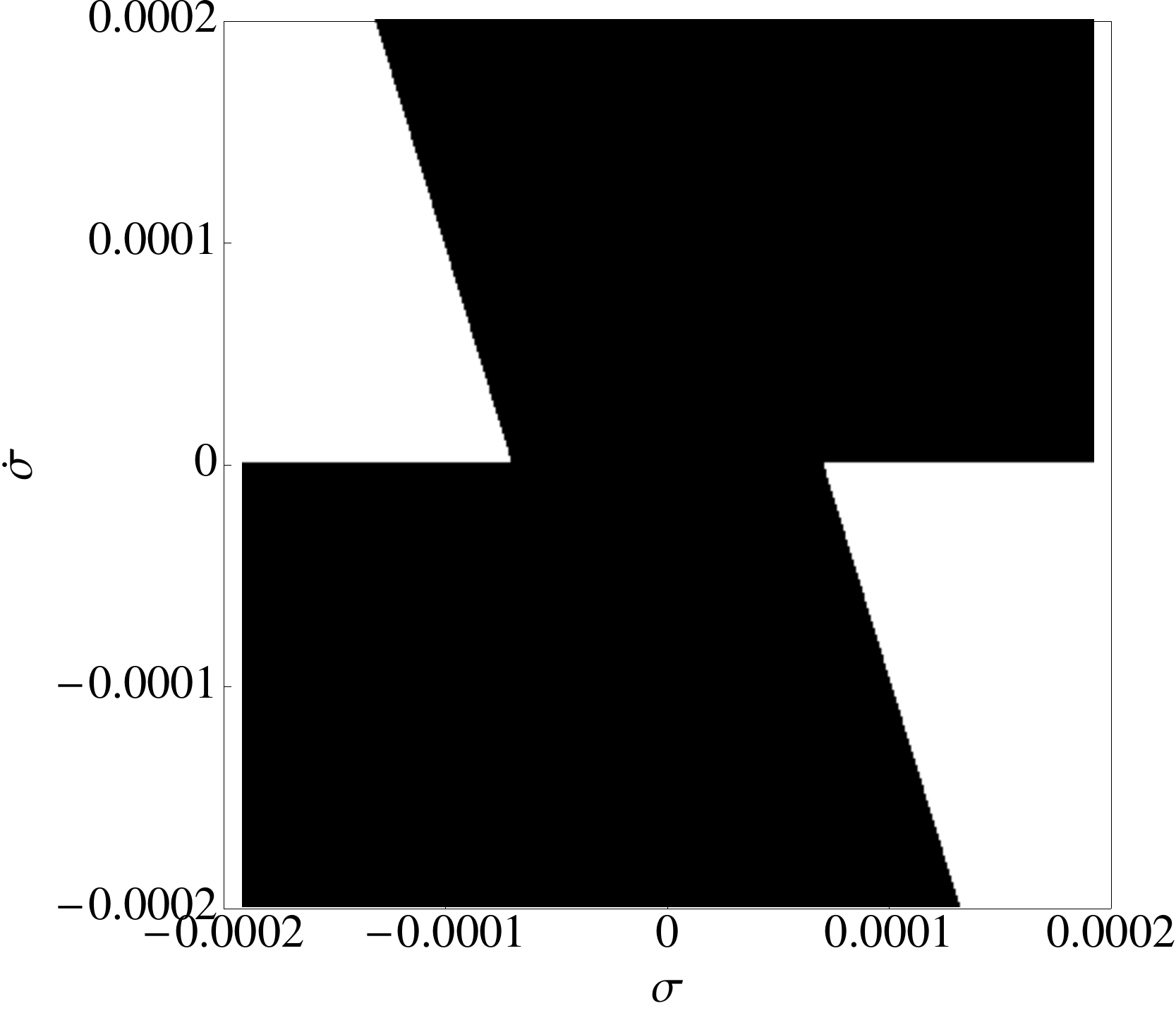}

\caption{The basins of attraction in the $\left(\sigma,\dot{\sigma}\right)$
plane for the closed FRW model. The slice is through $a=0.9$ and
$\dot{a}$ is fixed by the Hamiltonian constraint. We chose the parameters
$\lambda=16$ and $E_{0}=0.499\,999\,997\,5$. The black regions correspond
to collapse and the white regions inflation.\label{fig:fractal_sigma_frw}}
\end{figure}

\begin{figure}
\includegraphics[scale=0.5]{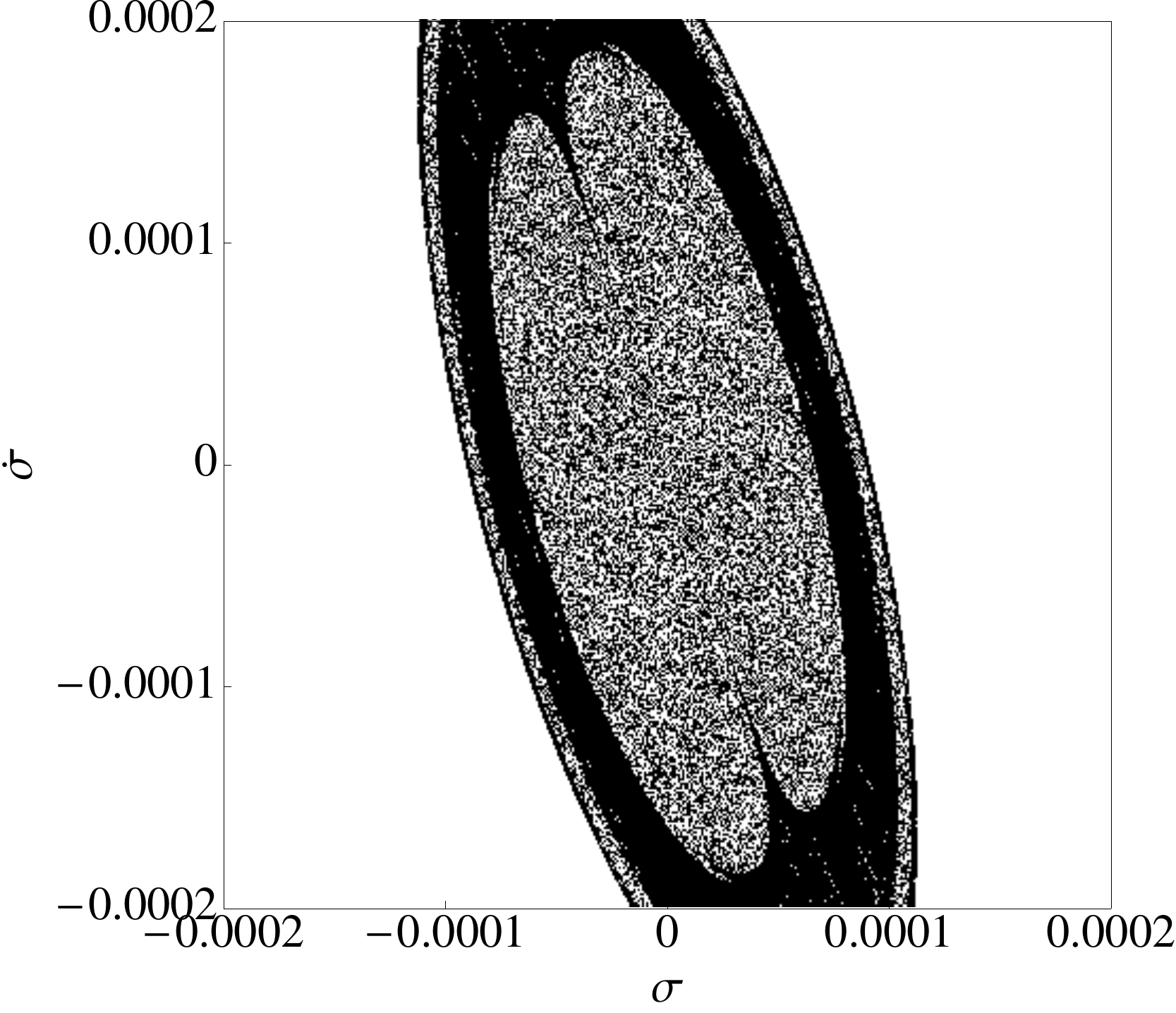}

\caption{The basins of attraction in the $\left(\sigma,\dot{\sigma}\right)$
plane for the axisymmetric Bianchi IX model. The slice is through
$a=b=0.9$, $\dot{a}=0.051\,818\,772\,5$, and $\dot{b}$ is fixed
by the Hamiltonian constraint. The other parameters are the same as
in Fig. \ref{fig:fractal_sigma_frw}.\label{fig:fractal_sigma_bianchi_IX}}
\end{figure}

\begin{figure}
\includegraphics[scale=0.5]{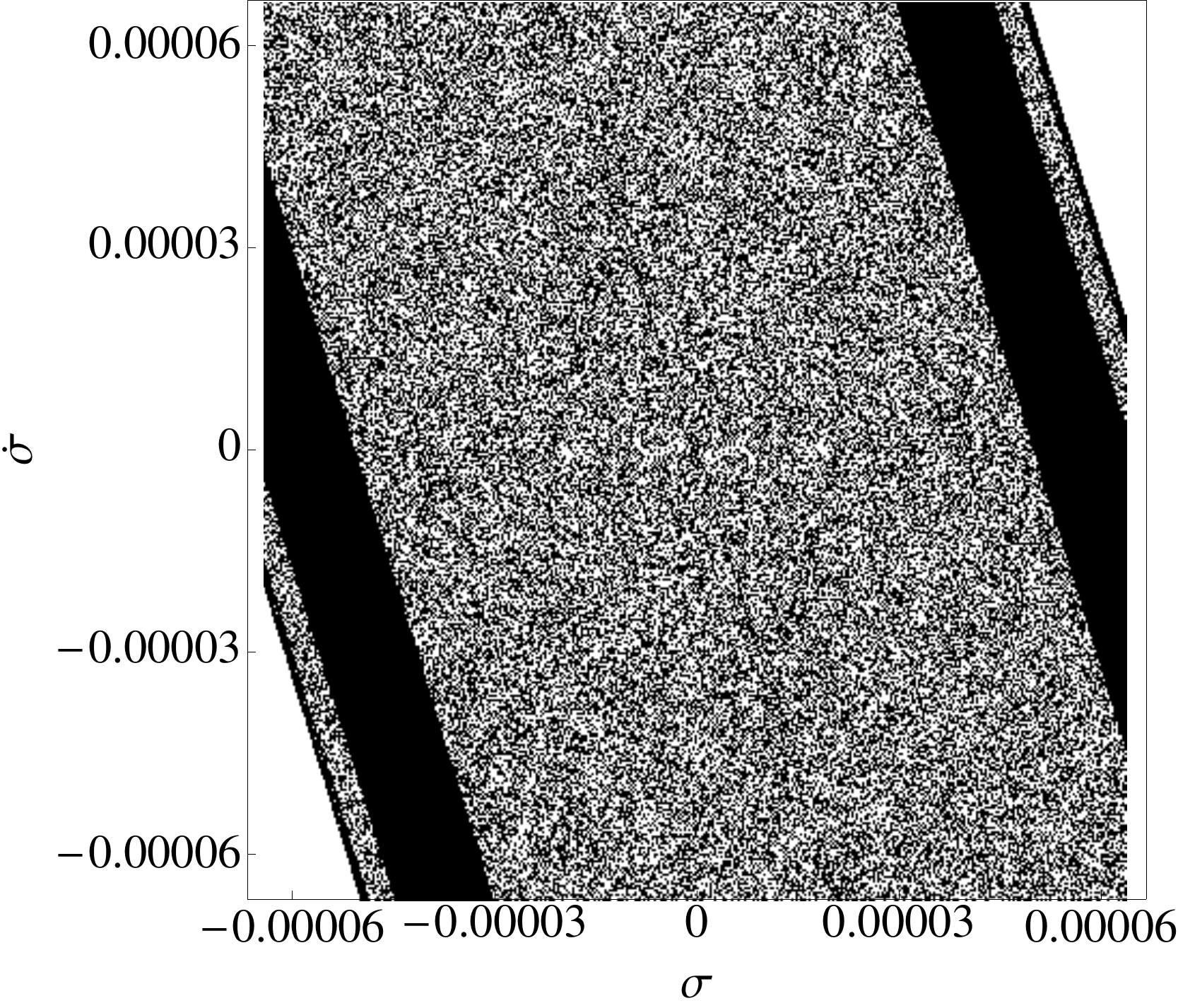}

\caption{A magnified view of Fig. \ref{fig:fractal_sigma_bianchi_IX}.\label{fig:fractal_sigma_bianchi_IX_magnified}}
\end{figure}

We now proceed to find a set of initial values that lead to an orbit
approaching the de Sitter attractor. Such a set is called a basin
of attraction. Here we use the standard method and do a pixel-by-pixel
computation of a $400\times400$ grid. Since we are dealing with basins
embedded in a six-dimensional phase space, we are forced to consider
lower dimensional slices, and here we choose the $\left(\sigma,\dot{\sigma}\right)$
plane although similar fractal basins of attraction can be obtained
for other slices in the phase space. As shown in Fig. \ref{fig:fractal_sigma_frw},
in the closed FRW universe, we have a sharply divided separatrix even
if we have a contribution from the Gauss-Bonnet term. When the metric
becomes anisotropic, the basins of attraction become highly fractal
as shown in Fig. \ref{fig:fractal_sigma_bianchi_IX}. A magnification
of the inner region is shown in Fig. \ref{fig:fractal_sigma_bianchi_IX_magnified}
and reveals self-similar fractal structure. Note that the black regions
correspond to orbits that collapse to the big crunch and the white
regions to orbits that expand to the de Sitter space. 

\begin{figure}
\includegraphics[scale=0.5]{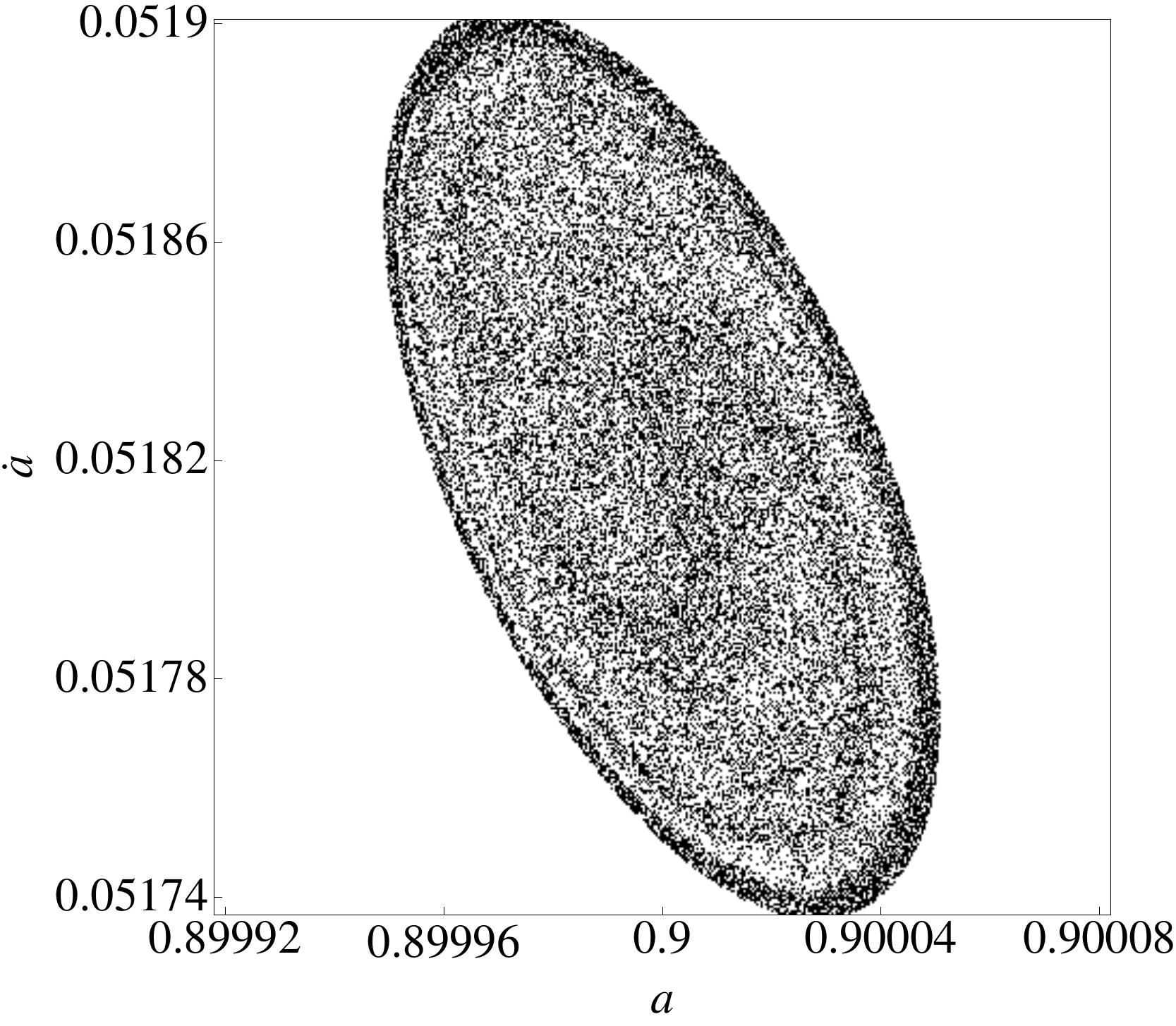}

\caption{The basins of attraction in the $\left(a,\dot{a}\right)$ plane. \label{fig:fractal_a}}
\end{figure}

The existence of fractal structures is not restricted to one particular
plane and can be seen in other slices in the six-dimensional phase
space. Fig. \ref{fig:fractal_a} shows that similar fractal structures
can also be seen in the $\left(a,\dot{a}\right)$ plane for universes
with parameters and conditions identical to those used in Fig. \ref{fig:fractal_sigma_bianchi_IX}.
These fractal structures in the phase space cannot be removed by coordinate
transformation and thus provide invariant characterization of chaos
in our model.

\section{Final Remarks}

In this paper, we have studied the dynamics of the closed FRW models
and Bianchi type-IX models in a ``string-inspired'' modified gravity,
which may provide a description of pre-inflationary stages of the
Universe after the Planck era. Higher curvature terms arise as the
next-to leading terms in the superstring effective action, or from
the renormalization of the stress tensor. We have included such effects
in the form of a Gauss-Bonnet curvature term non-minimally coupled
to a scalar field. The main features of our model are a positive cosmological
constant and a perfect fluid, which produce a saddle point in the
phase space. Due to the presence of the saddle point, an orbit starting
in the neighborhood of the separatrix has two asymptotic possibilities,
the de Sitter inflation and the big crunch collapse. In the closed
FRW model, the dynamics near this critical point is reduced to simple
two-dimensional hyperbolic motion of the Bogdanov-Takens bifurcation.
Consequently, the closed FRW model is not chaotic. In the anisotropic
Bianchi type-IX model, we restricted our attention to the axisymmetric
case. The introduction of anisotropy produces unstable periodic orbits,
which interact with simple hyperbolic motion of the planar Bogdanov-Takens
bifurcation to produce chaotic dynamics.

Our results extend the work of de Oliveira, Soares, and Stuchi in
\cite{deOliveira1997chaos}, who describe chaotic exit to inflation
for the axisymmetric Bianchi IX universe in general relativity. In
subsequent papers \cite{barguine2001homoclinic,de2002homoclinic,soares2005homoclinic}
the authors discuss the existence of ``homoclinic chaos'' in general
relativity. Our model does not have such homoclinic orbits because
of nonlinearity due to the coupling of Gauss-Bonnet curvature and
the scalar field. Furthermore, extending a cosmological model to the
$a<0$ region of the phase space is physically unreasonable as pointed
out in \cite{heinzle2006homoclinic}.

In the model considered in \cite{deOliveira1997chaos}, it is possible
to find an energy $E_{0}$ such that a small perturbation to a point
on the separatrix is a chaotic set. As we have shown in this paper,
this is also true when we include a stringy correction to the Einstein-Hilbert
action. We have shown that a small fluctuation in initial conditions
leads to indeterminate outcome between collapse or inflation. Furthermore,
we found some numerical evidence of fractal structures in the basins
of attraction. The fractal basins of attraction, together with topology
of cylinders near the critical point, are invariant characterization
of chaos in our model.
\begin{acknowledgments}
We acknowledge helpful conversations with Phillial Oh. This work was
supported in part by the National Research Foundation of Korea Grant-in-Aid
for Scientific Research No. 2012-007575 (S.K.) and the Center for
Quantum Spacetime (CQUeST) of Sogang University.
\end{acknowledgments}


\begin{thebibliography}{24}%
\makeatletter
\providecommand \@ifxundefined [1]{%
 \@ifx{#1\undefined}
}%
\providecommand \@ifnum [1]{%
 \ifnum #1\expandafter \@firstoftwo
 \else \expandafter \@secondoftwo
 \fi
}%
\providecommand \@ifx [1]{%
 \ifx #1\expandafter \@firstoftwo
 \else \expandafter \@secondoftwo
 \fi
}%
\providecommand \natexlab [1]{#1}%
\providecommand \enquote  [1]{``#1''}%
\providecommand \bibnamefont  [1]{#1}%
\providecommand \bibfnamefont [1]{#1}%
\providecommand \citenamefont [1]{#1}%
\providecommand \href@noop [0]{\@secondoftwo}%
\providecommand \href [0]{\begingroup \@sanitize@url \@href}%
\providecommand \@href[1]{\@@startlink{#1}\@@href}%
\providecommand \@@href[1]{\endgroup#1\@@endlink}%
\providecommand \@sanitize@url [0]{\catcode `\\12\catcode `\$12\catcode
  `\&12\catcode `\#12\catcode `\^12\catcode `\_12\catcode `\%12\relax}%
\providecommand \@@startlink[1]{}%
\providecommand \@@endlink[0]{}%
\providecommand \url  [0]{\begingroup\@sanitize@url \@url }%
\providecommand \@url [1]{\endgroup\@href {#1}{\urlprefix }}%
\providecommand \urlprefix  [0]{URL }%
\providecommand \Eprint [0]{\href }%
\providecommand \doibase [0]{http://dx.doi.org/}%
\providecommand \selectlanguage [0]{\@gobble}%
\providecommand \bibinfo  [0]{\@secondoftwo}%
\providecommand \bibfield  [0]{\@secondoftwo}%
\providecommand \translation [1]{[#1]}%
\providecommand \BibitemOpen [0]{}%
\providecommand \bibitemStop [0]{}%
\providecommand \bibitemNoStop [0]{.\EOS\space}%
\providecommand \EOS [0]{\spacefactor3000\relax}%
\providecommand \BibitemShut  [1]{\csname bibitem#1\endcsname}%
\let\auto@bib@innerbib\@empty
\bibitem [{\citenamefont {Belinskii}\ \emph {et~al.}(1969)\citenamefont
  {Belinskii}, \citenamefont {Lifshitz},\ and\ \citenamefont
  {Khalatnikov}}]{BKL1969}%
  \BibitemOpen
  \bibfield  {author} {\bibinfo {author} {\bibfnamefont {V.}~\bibnamefont
  {Belinskii}}, \bibinfo {author} {\bibfnamefont {E.}~\bibnamefont {Lifshitz}},
  \ and\ \bibinfo {author} {\bibfnamefont {I.}~\bibnamefont {Khalatnikov}},\
  }\href@noop {} {\bibfield  {journal} {\bibinfo  {journal} {Usp. Fiz. Nauk.}\
  }\textbf {\bibinfo {volume} {56}},\ \bibinfo {pages} {1700} (\bibinfo {year}
  {1969})},\ \bibinfo {note} {[Sov. Phys. JETP 29(5):911-917
  (1969)]}\BibitemShut {NoStop}%
\bibitem [{\citenamefont {Misner}(1969)}]{misner1969mixmaster}%
  \BibitemOpen
  \bibfield  {author} {\bibinfo {author} {\bibfnamefont {C.}~\bibnamefont
  {Misner}},\ }\href@noop {} {\bibfield  {journal} {\bibinfo  {journal}
  {Physical Review Letters}\ }\textbf {\bibinfo {volume} {22}},\ \bibinfo
  {pages} {1071} (\bibinfo {year} {1969})}\BibitemShut {NoStop}%
\bibitem [{\citenamefont {Francisco}\ and\ \citenamefont
  {Matsas}(1988)}]{francisco1988qualitative}%
  \BibitemOpen
  \bibfield  {author} {\bibinfo {author} {\bibfnamefont {G.}~\bibnamefont
  {Francisco}}\ and\ \bibinfo {author} {\bibfnamefont {G.}~\bibnamefont
  {Matsas}},\ }\href@noop {} {\bibfield  {journal} {\bibinfo  {journal}
  {General relativity and gravitation}\ }\textbf {\bibinfo {volume} {20}},\
  \bibinfo {pages} {1047} (\bibinfo {year} {1988})}\BibitemShut {NoStop}%
\bibitem [{\citenamefont {Burd}\ \emph {et~al.}(1991)\citenamefont {Burd},
  \citenamefont {Buric},\ and\ \citenamefont {Tavakol}}]{burd1991chaos}%
  \BibitemOpen
  \bibfield  {author} {\bibinfo {author} {\bibfnamefont {A.}~\bibnamefont
  {Burd}}, \bibinfo {author} {\bibfnamefont {N.}~\bibnamefont {Buric}}, \ and\
  \bibinfo {author} {\bibfnamefont {R.}~\bibnamefont {Tavakol}},\ }\href@noop
  {} {\bibfield  {journal} {\bibinfo  {journal} {Classical and Quantum
  Gravity}\ }\textbf {\bibinfo {volume} {8}},\ \bibinfo {pages} {123} (\bibinfo
  {year} {1991})}\BibitemShut {NoStop}%
\bibitem [{\citenamefont {Berger}(1991)}]{berger1991comments}%
  \BibitemOpen
  \bibfield  {author} {\bibinfo {author} {\bibfnamefont {B.}~\bibnamefont
  {Berger}},\ }\href@noop {} {\bibfield  {journal} {\bibinfo  {journal}
  {General relativity and gravitation}\ }\textbf {\bibinfo {volume} {23}},\
  \bibinfo {pages} {1385} (\bibinfo {year} {1991})}\BibitemShut {NoStop}%
\bibitem [{\citenamefont {Rugh}(1990)}]{Rugh1990}%
  \BibitemOpen
  \bibfield  {author} {\bibinfo {author} {\bibfnamefont {S.}~\bibnamefont
  {Rugh}},\ }\href@noop {} {Ph.D. thesis},\ \bibinfo  {school} {The Niels Bohr
  Institute} (\bibinfo {year} {1990})\BibitemShut {NoStop}%
\bibitem [{\citenamefont {Cornish}\ and\ \citenamefont
  {Levin}(1996)}]{cornish1996chaos}%
  \BibitemOpen
  \bibfield  {author} {\bibinfo {author} {\bibfnamefont {N.}~\bibnamefont
  {Cornish}}\ and\ \bibinfo {author} {\bibfnamefont {J.}~\bibnamefont
  {Levin}},\ }\href@noop {} {\bibfield  {journal} {\bibinfo  {journal}
  {Physical Review D}\ }\textbf {\bibinfo {volume} {53}},\ \bibinfo {pages}
  {3022} (\bibinfo {year} {1996})}\BibitemShut {NoStop}%
\bibitem [{\citenamefont {De~Oliveira}\ \emph {et~al.}(1997)\citenamefont
  {De~Oliveira}, \citenamefont {Soares},\ and\ \citenamefont
  {Stuchi}}]{deOliveira1997chaos}%
  \BibitemOpen
  \bibfield  {author} {\bibinfo {author} {\bibfnamefont {H.}~\bibnamefont
  {De~Oliveira}}, \bibinfo {author} {\bibfnamefont {I.}~\bibnamefont {Soares}},
  \ and\ \bibinfo {author} {\bibfnamefont {T.}~\bibnamefont {Stuchi}},\
  }\href@noop {} {\bibfield  {journal} {\bibinfo  {journal} {Physical Review
  D}\ }\textbf {\bibinfo {volume} {56}},\ \bibinfo {pages} {730} (\bibinfo
  {year} {1997})}\BibitemShut {NoStop}%
\bibitem [{\citenamefont {Brandenberger}\ and\ \citenamefont
  {Martin}(2012)}]{Brandenberger:2012aj}%
  \BibitemOpen
  \bibfield  {author} {\bibinfo {author} {\bibfnamefont {R.}~\bibnamefont
  {Brandenberger}}\ and\ \bibinfo {author} {\bibfnamefont {J.}~\bibnamefont
  {Martin}},\ }\href@noop {} {\  (\bibinfo {year} {2012})},\ \Eprint
  {http://arxiv.org/abs/1211.6753} {arXiv:1211.6753 [astro-ph.CO]} \BibitemShut
  {NoStop}%
\bibitem [{\citenamefont {Cotsakis}\ \emph {et~al.}(1993)\citenamefont
  {Cotsakis}, \citenamefont {Demaret}, \citenamefont {De~Rop},\ and\
  \citenamefont {Querella}}]{cotsakis1993mixmaster}%
  \BibitemOpen
  \bibfield  {author} {\bibinfo {author} {\bibfnamefont {S.}~\bibnamefont
  {Cotsakis}}, \bibinfo {author} {\bibfnamefont {J.}~\bibnamefont {Demaret}},
  \bibinfo {author} {\bibfnamefont {Y.}~\bibnamefont {De~Rop}}, \ and\ \bibinfo
  {author} {\bibfnamefont {L.}~\bibnamefont {Querella}},\ }\href@noop {}
  {\bibfield  {journal} {\bibinfo  {journal} {Physical Review D}\ }\textbf
  {\bibinfo {volume} {48}},\ \bibinfo {pages} {4595} (\bibinfo {year}
  {1993})}\BibitemShut {NoStop}%
\bibitem [{\citenamefont {Yajima}\ \emph {et~al.}(2000)\citenamefont {Yajima},
  \citenamefont {Maeda},\ and\ \citenamefont {Ohkubo}}]{yajima2000generality}%
  \BibitemOpen
  \bibfield  {author} {\bibinfo {author} {\bibfnamefont {H.}~\bibnamefont
  {Yajima}}, \bibinfo {author} {\bibfnamefont {K.}~\bibnamefont {Maeda}}, \
  and\ \bibinfo {author} {\bibfnamefont {H.}~\bibnamefont {Ohkubo}},\
  }\href@noop {} {\bibfield  {journal} {\bibinfo  {journal} {Physical Review
  D}\ }\textbf {\bibinfo {volume} {62}},\ \bibinfo {pages} {024020} (\bibinfo
  {year} {2000})}\BibitemShut {NoStop}%
\bibitem [{\citenamefont {Antoniadis}\ \emph {et~al.}(1994)\citenamefont
  {Antoniadis}, \citenamefont {Gava}, \citenamefont {Narain},\ and\
  \citenamefont {Taylor}}]{antoniadis1994topological}%
  \BibitemOpen
  \bibfield  {author} {\bibinfo {author} {\bibfnamefont {I.}~\bibnamefont
  {Antoniadis}}, \bibinfo {author} {\bibfnamefont {E.}~\bibnamefont {Gava}},
  \bibinfo {author} {\bibfnamefont {K.}~\bibnamefont {Narain}}, \ and\ \bibinfo
  {author} {\bibfnamefont {T.}~\bibnamefont {Taylor}},\ }\href@noop {}
  {\bibfield  {journal} {\bibinfo  {journal} {Nuclear Physics B}\ }\textbf
  {\bibinfo {volume} {413}},\ \bibinfo {pages} {162} (\bibinfo {year}
  {1994})}\BibitemShut {NoStop}%
\bibitem [{\citenamefont {Rizos}\ and\ \citenamefont
  {Tamvakis}(1994)}]{rizos1994existence}%
  \BibitemOpen
  \bibfield  {author} {\bibinfo {author} {\bibfnamefont {J.}~\bibnamefont
  {Rizos}}\ and\ \bibinfo {author} {\bibfnamefont {K.}~\bibnamefont
  {Tamvakis}},\ }\href@noop {} {\bibfield  {journal} {\bibinfo  {journal}
  {Physics Letters B}\ }\textbf {\bibinfo {volume} {326}},\ \bibinfo {pages}
  {57} (\bibinfo {year} {1994})}\BibitemShut {NoStop}%
\bibitem [{\citenamefont {Easther}\ and\ \citenamefont
  {Maeda}(1996)}]{easther1996one}%
  \BibitemOpen
  \bibfield  {author} {\bibinfo {author} {\bibfnamefont {R.}~\bibnamefont
  {Easther}}\ and\ \bibinfo {author} {\bibfnamefont {K.}~\bibnamefont
  {Maeda}},\ }\href@noop {} {\bibfield  {journal} {\bibinfo  {journal}
  {Physical Review D}\ }\textbf {\bibinfo {volume} {54}},\ \bibinfo {pages}
  {7252} (\bibinfo {year} {1996})}\BibitemShut {NoStop}%
\bibitem [{\citenamefont {Kawai}\ \emph {et~al.}(1998)\citenamefont {Kawai},
  \citenamefont {Sakagami},\ and\ \citenamefont {Soda}}]{kawai1998instability}%
  \BibitemOpen
  \bibfield  {author} {\bibinfo {author} {\bibfnamefont {S.}~\bibnamefont
  {Kawai}}, \bibinfo {author} {\bibfnamefont {M.}~\bibnamefont {Sakagami}}, \
  and\ \bibinfo {author} {\bibfnamefont {J.}~\bibnamefont {Soda}},\ }\href@noop
  {} {\bibfield  {journal} {\bibinfo  {journal} {Physics Letters B}\ }\textbf
  {\bibinfo {volume} {437}},\ \bibinfo {pages} {284} (\bibinfo {year}
  {1998})}\BibitemShut {NoStop}%
\bibitem [{\citenamefont {Kawai}\ and\ \citenamefont
  {Soda}(1999)}]{kawai1999nonsingular}%
  \BibitemOpen
  \bibfield  {author} {\bibinfo {author} {\bibfnamefont {S.}~\bibnamefont
  {Kawai}}\ and\ \bibinfo {author} {\bibfnamefont {J.}~\bibnamefont {Soda}},\
  }\href@noop {} {\bibfield  {journal} {\bibinfo  {journal} {Physical Review
  D}\ }\textbf {\bibinfo {volume} {59}},\ \bibinfo {pages} {063506} (\bibinfo
  {year} {1999})}\BibitemShut {NoStop}%
\bibitem [{\citenamefont {Antoniadis}\ \emph {et~al.}(1992)\citenamefont
  {Antoniadis}, \citenamefont {Gava},\ and\ \citenamefont
  {Narain}}]{antoniadis1992moduli}%
  \BibitemOpen
  \bibfield  {author} {\bibinfo {author} {\bibfnamefont {I.}~\bibnamefont
  {Antoniadis}}, \bibinfo {author} {\bibfnamefont {E.}~\bibnamefont {Gava}}, \
  and\ \bibinfo {author} {\bibfnamefont {K.}~\bibnamefont {Narain}},\
  }\href@noop {} {\bibfield  {journal} {\bibinfo  {journal} {Nuclear Physics
  B}\ }\textbf {\bibinfo {volume} {383}},\ \bibinfo {pages} {93} (\bibinfo
  {year} {1992})}\BibitemShut {NoStop}%
\bibitem [{\citenamefont {Barguine}\ \emph {et~al.}(2001)\citenamefont
  {Barguine}, \citenamefont {de~Oliveira}, \citenamefont {Soares},\ and\
  \citenamefont {Tonini}}]{barguine2001homoclinic}%
  \BibitemOpen
  \bibfield  {author} {\bibinfo {author} {\bibfnamefont {R.}~\bibnamefont
  {Barguine}}, \bibinfo {author} {\bibfnamefont {H.}~\bibnamefont
  {de~Oliveira}}, \bibinfo {author} {\bibfnamefont {I.}~\bibnamefont {Soares}},
  \ and\ \bibinfo {author} {\bibfnamefont {E.}~\bibnamefont {Tonini}},\
  }\href@noop {} {\bibfield  {journal} {\bibinfo  {journal} {Physical Review
  D}\ }\textbf {\bibinfo {volume} {63}},\ \bibinfo {pages} {063502} (\bibinfo
  {year} {2001})}\BibitemShut {NoStop}%
\bibitem [{\citenamefont {Wiggins}(2003)}]{wiggins2003introduction}%
  \BibitemOpen
  \bibfield  {author} {\bibinfo {author} {\bibfnamefont {S.}~\bibnamefont
  {Wiggins}},\ }\href@noop {} {\emph {\bibinfo {title} {Introduction to applied
  nonlinear dynamical systems and chaos}}},\ Vol.~\bibinfo {volume} {2}\
  (\bibinfo  {publisher} {Springer},\ \bibinfo {year} {2003})\BibitemShut
  {NoStop}%
\bibitem [{\citenamefont {Bogdanov}(1975)}]{bogdanov1975versal}%
  \BibitemOpen
  \bibfield  {author} {\bibinfo {author} {\bibfnamefont {R.}~\bibnamefont
  {Bogdanov}},\ }\href@noop {} {\bibfield  {journal} {\bibinfo  {journal}
  {Functional analysis and its applications}\ }\textbf {\bibinfo {volume}
  {9}},\ \bibinfo {pages} {144} (\bibinfo {year} {1975})}\BibitemShut {NoStop}%
\bibitem [{\citenamefont {Arnold}\ and\ \citenamefont
  {Levi}(1988)}]{arnold1988geometricalmethods}%
  \BibitemOpen
  \bibfield  {author} {\bibinfo {author} {\bibfnamefont {V.}~\bibnamefont
  {Arnold}}\ and\ \bibinfo {author} {\bibfnamefont {M.}~\bibnamefont {Levi}},\
  }\href@noop {} {\emph {\bibinfo {title} {Geometrical Methods in the Theory of
  Ordinary Differential Equations}}}\ (\bibinfo  {publisher}
  {Springer-Verlag},\ \bibinfo {year} {1988})\BibitemShut {NoStop}%
\bibitem [{\citenamefont {De~Oliveira}\ \emph {et~al.}(2002)\citenamefont
  {De~Oliveira}, \citenamefont {de~Almeida}, \citenamefont {Soares},\ and\
  \citenamefont {Tonini}}]{de2002homoclinic}%
  \BibitemOpen
  \bibfield  {author} {\bibinfo {author} {\bibfnamefont {H.}~\bibnamefont
  {De~Oliveira}}, \bibinfo {author} {\bibfnamefont {A.}~\bibnamefont
  {de~Almeida}}, \bibinfo {author} {\bibfnamefont {I.}~\bibnamefont {Soares}},
  \ and\ \bibinfo {author} {\bibfnamefont {E.}~\bibnamefont {Tonini}},\
  }\href@noop {} {\bibfield  {journal} {\bibinfo  {journal} {Physical Review
  D}\ }\textbf {\bibinfo {volume} {65}},\ \bibinfo {pages} {083511} (\bibinfo
  {year} {2002})}\BibitemShut {NoStop}%
\bibitem [{\citenamefont {Soares}\ and\ \citenamefont
  {Stuchi}(2005)}]{soares2005homoclinic}%
  \BibitemOpen
  \bibfield  {author} {\bibinfo {author} {\bibfnamefont {I.}~\bibnamefont
  {Soares}}\ and\ \bibinfo {author} {\bibfnamefont {T.}~\bibnamefont
  {Stuchi}},\ }\href@noop {} {\bibfield  {journal} {\bibinfo  {journal}
  {Physical Review D}\ }\textbf {\bibinfo {volume} {72}},\ \bibinfo {pages}
  {083516} (\bibinfo {year} {2005})}\BibitemShut {NoStop}%
\bibitem [{\citenamefont {Heinzle}\ \emph {et~al.}(2006)\citenamefont
  {Heinzle}, \citenamefont {R{\"o}hr},\ and\ \citenamefont
  {Uggla}}]{heinzle2006homoclinic}%
  \BibitemOpen
  \bibfield  {author} {\bibinfo {author} {\bibfnamefont {J.}~\bibnamefont
  {Heinzle}}, \bibinfo {author} {\bibfnamefont {N.}~\bibnamefont {R{\"o}hr}}, \
  and\ \bibinfo {author} {\bibfnamefont {C.}~\bibnamefont {Uggla}},\
  }\href@noop {} {\bibfield  {journal} {\bibinfo  {journal} {Physical Review
  D}\ }\textbf {\bibinfo {volume} {74}},\ \bibinfo {pages} {061502} (\bibinfo
  {year} {2006})}\BibitemShut {NoStop}%
\end{thebibliography}

%

\end{document}